\documentclass[aps,prb,twocolumn,showpacs,superscriptaddress,floatfix,
amsmath,amssymb]{revtex4}
\usepackage{graphicx}
\newcommand{\beq}{\begin{equation}}
\newcommand{\eeq}{\end{equation}}
\newcommand{\bea}{\begin{eqnarray}}
\newcommand{\eea}{\end{eqnarray}}
\begin{document}
\title{On R\'enyi entropies characterizing the shape and the extension 
of the phase space representation of quantum wave functions in 
disordered systems}
\author{Imre Varga}
\affiliation{Elm\'eleti Fizika Tansz\'ek,
Budapesti M\H uszaki \'es Gazdas\'agtudom\'anyi Egyetem,
H-1521 Budapest, Hungary}
\affiliation{Fachbereich Physik, Philipps--Universit\"at Marburg,
Renthof 6, D-35032 Marburg, Germany}
\author{J\'anos Pipek}
\affiliation{Elm\'eleti Fizika Tansz\'ek,
Budapesti M\H uszaki \'es Gazdas\'agtudom\'anyi Egyetem,
H-1521 Budapest, Hungary}
\date{\today}
\begin{abstract}
We discuss some properties of the generalized entropies, called
R\'enyi entropies and their application to the case of continuous
distributions. In particular it is shown that these measures of
complexity can be divergent, however, their differences are free from
these divergences thus enabling them to be good candidates for the
description of the extension and the shape of continuous distributions. 
We apply this formalism to the projection of wave functions onto 
the coherent state basis, i.e. to the Husimi representation. We also 
show how the localization properties of the Husimi distribution on 
average can be reconstructed from its marginal distributions that are 
calculated in position and momentum space in the case when the phase 
space has no structure, i.e. no classical limit can be defined. 
Numerical simulations on a one dimensional disordered system corroborate 
our expectations.
\end{abstract}
\maketitle{}
\section{Introduction}

The measure of the extension of phase space distribution of quantum
states tells us important information on the degree of ergodicity and
at the same time the degree of localization. These information are
directly connected to the caoticity of the underlying classical
dynamics if the latter is meaningful. In the ergodic regime trajectories 
visit every corner of phase space hence the quantum states associated 
to such orbits are expected to be extended. On the other hand regular 
islands trap classical trajectories and the corresponding states are 
localized. Therefore the extension properties of the eigenstates directly 
reflect the nature of classical dynamics. For this purpose the Shannon 
entropy or information content has been used widely as a measure of 
complexity of quantum states. This and other generalized entropic measures 
have been invoked by \.Zyczkowsky in Ref. \onlinecite{karol90} as measures
of chaotic signatures. In subsequent work \cite{karolphe} this idea has
been elaborated further and a direct correspondence between the 
complexity of quantum states and the underlying dynamics has been
demonstrated in particular by projecting the quantum states onto a
coherent state basis. Recently other works have showed the renewed
interest in this field \cite{AH,Korsch,Karol01,sugita}. We have to 
emphasize, however, that even systems without classical limit, e.g. 
disordered systems, have been involved in such phase space studies 
\cite{dietmar,wobst,heis}. This latter topic is the main motivation of
our present work as well.

In the present work we give further arguments in favor of the
application of generalized entropies, the R\'enyi entropies, for the
characterization of quantum phase space distributions, however, we
point out some problems in connection to the calculation of these 
parameters for continuous distributions. As a remedy for these problems
we show that the differences of R\'enyi entropies are on the other hand
free from these anomalies.

Furthermore we will show that indeed these entropic measures give
important information concerning the localization properties of phase
space distributions especially the Husimi distribution. We will also 
give arguments and numerical proofs that in fact in the case of wave
functions of disordered systems there is no need to calculate 
the Husimi distributions themselves especially because the average
localization properties of the Husimi functions of a set of states
can be obtained from the average properties of the marginal distributions
of the Husimi functions.

Using different techniques similar 
results have been obtained for a particular quantity, the participation
ratio in Refs. \onlinecite{sugita,wobst}. Our approach, however, is
more general.

In the next section we introduce the basic ideas and tools that have
been widely used in the literature, namely the participation number
(ratio) and the Shannon entropy for the case of discrete distributions. 
We also show that these quantities give roughly the same information, 
however, using these parameters a new quantity, the structural entropy, 
can be defined that contains important information concerning the 
shape of a distribution. It is also shown that these parameters are 
nothing else but some special cases of the differences of R\'enyi 
entropies. Section II contains merely the revision of what has been 
published before and we conclude this section analyzing the problem 
of continuous distributions and showing that the above mentioned 
differences of R\'enyi entropies are free from the divergences.
In Section III we elaborate the appropriate
generalization of these parameters for continuous distributions. In
Section IV we introduce the Husimi representation of quantum states
and describe some of its properties. In Section V R\'enyi entropies
are applied for Husimi distributions and it is shown that the
properties of its marginal distributions already give a qualitative
picture that for special cases may quantitatively be correct, 
as well. In Sect. VI numerical 
simulations for the one dimensional Anderson model provide important 
verification of the results presented in Sect. V. Finally some
concluding remark are left for Sect. VII.

\section{Basic ideas}
The extension of a discrete distribution of a state is often 
characterized by its entropy, $S$, or by its second moment, $D$. 
Both of these quantities, i.e. $\exp (S)$ and $D$ practically measure 
the same thing, namely the number of amplitudes that mainly contribute 
to the expansion of the state over a suitable basis.

Let us consider a wave function $\Psi$ that is represented by its
expansion over a complete basis set on a finite grid of $N$ states $\phi_i$
\beq
\Psi=\sum_{i=1}^Nc_i\phi_i\qquad\mbox{with}\qquad\sum_{i=1}^N|c_i|^2=1
\eeq
Note that each of the coefficients, $Q_i=|c_i|^2$, obey the condition
\beq
0\leq Q_i\leq 1, 
\label{01}
\eeq
and they sum up to unity.
Then the usual definitions of participation number, $D$, and entropy,
$S$ are
\beq
D^{-1}=\sum_{i=1}^NQ_i^2
\qquad\mbox{and}\qquad
S=-\sum_{i=1}^NQ_i\ln Q_i.
\label{D&S}
\eeq
The parameter $D$ tells us how many of the numbers $Q_i$ are
significantly larger than zero. For example if only one of them is
unity and the rest is zero, then $D=1$. Otherwise if $Q_i=1/N$
homogeneously, then we get $D=N$. Similar properties can be shown 
to hold for $\exp S$, therefore it is easy to show that the two 
quantities provide roughly the same information
\beq
S\approx \ln D,
\label{seqd}
\eeq
i.e. both, $S$ and $\ln D$ describe the extension of the discrete
distribution. That is the reason for calling $D$ as the
number of principal components. The close relation between $S$ and
$D$, Eq.~(\ref{seqd}), has often been overlooked and presented
\cite{manyb,Felix} as an interesting similarity.
However, as it has been demonstrated in Ref. \onlinecite{PV92} and
applied in several studies later \cite{PV92,other2,quchem,dqp,spst},
the difference
\beq
S_{str}=S-\ln D
\label{sstr}
\eeq
is a meaningful and most importantly a
nonnegative quantity that turns out to be very useful
in the characterization of the shape of the distribution of the
probabilities $Q_i$. That is the reason why it has been termed as
structural entropy of a distribution. Moreover the value of the
participation number normalized to the number of available components 
is an important partner quantity
\beq
q=\frac{D}{N},
\label{ff}
\eeq
which has been termed as the participation ratio in the literature.
These two quantities satisfy the following inequalities \cite{PV92}
\begin{subequations}
\bea
0<&q&\leq 1\label{q1}\label{ineq}\\
0\leq&S_{str}&\leq-\ln q\label{ines}.
\eea
\label{qsine}
\end{subequations}

Generalized entropies have been introduced by R\'enyi \cite{renyi} in
the form of
\beq
R_m=\frac{1}{1-m}\ln\sum_{i=1}^NQ_i^m,
\label{renyi1}
\eeq
that monotonously decreases for increasing $m$.
For the special cases of $m=0$, $1$, and $2$ we recover the total
number of components, the Shannon-entropy and the participation number 
\beq
R_0=\ln N\qquad
\lim_{m\to 1}R_m=S,\qquad
R_2=\ln D.
\eeq
Notice that the order $m$ in (\ref{renyi1}) is not necessarily integer.
We can readily realize that the parameters, (\ref{sstr}) and
(\ref{ff}), are nothing else but the differences \cite{other2} of the
special cases of $R_0$, $R_1$, and $R_2$, i.e.
\begin{subequations}
\bea
S_{str}&=&R_1-R_2\label{qrenyi}\\ 
-\ln q&=&R_0-R_2\label{srenyi}.
\eea
\label{sq_renyi}
\end{subequations}
A number of applications have been presented to date \cite{other2}
showing their diverse applicability starting from quantum chemistry
\cite{quchem} up to localization in disordered and quasiperiodic
systems \cite{dqp} up to the statistical analysis of spectra
\cite{spst}.

In the present work we are going to extend this formalism to
continuous distributions and show again that the differences of
R\'enyi entropies are good candidates for the characterization of
them.

The problem with a continuous distribution $p(x)$ is that even though
normalization requires
\beq
\int dx\,p(x)=1,
\label{norm}
\eeq
it is clear that $p(x)$ is a density of probabilities, therefore it is
the quantity $p(x)\Delta x$, the probability associated with the
interval $[x,x+\Delta x]$ that is restricted to the $[0,1]$ interval
and not the value of $p(x)$ itself. Hence even though we may always
expect $p(x)\geq 0$ the condition $p(x)<1$ is generally not fulfilled.

Then the obvious generalization of the R\'enyi entropies
(\ref{renyi1}) for normalized continuous distributions would read as
\beq
R_m=\frac{1}{1-m}\ln\int dx\,[p(x)]^m.
\label{renyi2}
\eeq
Hence the definitions of the participation number and the entropy
(\ref{D&S}) would look as
\bea
\label{DS}
D^{-1}&=&\int dx\,[p(x)]^2\label{DS_d}\\
S&=&-\int dx\,p(x)\ln p(x).\label{DS_s}
\eea
Let us apply this definition to a Gaussian distribution with zero mean
and variance $\sigma$
\beq
p(x)=\frac{1}{\sqrt{2\pi\sigma^2}}\exp\left(-\frac{x^2}{2\sigma^2}\right).
\label{gauss}
\eeq
The formulas derived using this $p(x)$ will be useful when applying
for the problem of Husimi distributions later due to the Gaussian
smearing contained in those phase space functions. Putting
Eq.~(\ref{gauss}) in Eq.~(\ref{renyi2}) we obtain for $m>0$
\beq
R_m(\sigma)=\frac{\ln\sqrt{m}}{m-1}+\ln (\sqrt{2\pi}\sigma),
\label{re1}
\eeq
which tells us that as $\sigma\to 0$ or $\sigma\to\infty$ the R\'enyi
entropies diverge. However, they do that uniformly, i.e. independently
of $m$, hence their differences remain finite. This is an important 
advantage of our formulation that will be elaborated further in the 
subsequent part. In particular in the case of the Gaussian (\ref{gauss}) 
we obtain $q=0$ and
\beq
S_{str}^{G}=R_1-R_2=\frac{1}{2}(1-\ln 2)=0.1534\dots
\label{ggg}
\eeq
This is the value of the structural entropy that describes a Gaussian 
distribution in one dimension.

Next let us consider the above Gaussian on a finite interval,
$-L/2\leq x\leq L/2$ and assume that beyond this interval $p(x)=0$. 
This construction allows us to study how the limit of
(\ref{re1}) or (\ref{ggg}) is approached as for fixed $\sigma$ 
the interval tends to infinity (or for a fixed $L$ the width
$\sigma\to 0$). To this end the normalization is taken over the finite
interval $[-L/2,L/2]$ so the distribution function (\ref{gauss})
should be modified as
\beq
p(x)=\frac{1}{\sqrt{2\pi\sigma^2}}\frac{e^{-x^2/2\sigma^2}}{\Phi(\xi/\sqrt{8})}
\label{ngauss}
\eeq
where $\Phi(x)$ denotes the error function and the scaling parameter 
$\xi=L/\sigma$ has been introduced. The R\'enyi entropies will depend 
on $L$ and $\sigma$ as 
\beq
R_m(L,\sigma)=R_m(\infty,\sigma)+
\frac{1}{m-1}
\ln\left[\frac {[\Phi\left(\xi/\sqrt{8}\right)]^m}
               {\Phi\left(\xi\sqrt{m/8}\right)}\right]
\label{rexi}
\eeq
where $R_m(\infty,\sigma)$ is given in Eq.~(\ref{re1}). Again the
$m$--independent $-\ln\sigma$ divergence appears. Turning to the 
special cases of $q$ and $S_{str}$ as deduced from $R_0$, $R_1$, 
and $R_2$ using Eqs. (\ref{sq_renyi}) as before we find that they 
are uniquely determined by $\xi$ and are free from this type of 
divergence. In particular since $R_0=\ln L$, 
\beq
q(\xi)=\frac{2\sqrt{\pi}}{\xi}
       \frac{[\Phi(\xi/\sqrt{8})]^2}{\Phi(\xi/2)}
\label{qxi}
\eeq
and
\beq
S_{str}(\xi)=S_{str}^{G}
             -\frac{\xi e^{-\xi^2/8}}{\sqrt{8\pi}\Phi(\xi/\sqrt{8})}
             +\ln\left[\frac{\Phi(\xi/2)}{\Phi(\xi/\sqrt{8})}\right],
\label{sxi}
\eeq
where $S_{str}^{G}$ is given in Eq.~(\ref{ggg}). The participation
ratio, $q(\xi)$, in the limit $\xi\to\infty$ ($L\to\infty$ for fixed
$\sigma$ or $\sigma\to 0$ for fixed $L$) tends to
zero as $q(\xi)\approx 2\sqrt{\pi}/\xi$ while $S_{str}(\xi)\to
S_{str}^G$.  On the other hand in the other
limit of $\xi\to 0$ ($\sigma\to\infty$ for fixed $L$ or $L\to 0$ for
fixed $\sigma$) we see that $q(\xi)\approx (1-\xi^4/720)$ and
$S_{str}(\xi)\approx\xi^4/1440$ therefore the relation 
$S_{str}\approx (1-q)/2$ is also fulfilled \cite{PV92}. We would like 
to emphasize that no divergences are found for parameters $q$ and 
$S_{str}$ and they show well defined behavior in either limit.

\section{Coarse graining}
Now let us turn to a more general investigation of our parameters. In
this section we provide a natural generalization of the calculation of
the parameters $q$ and $S_{str}$ for continuous distributions. 

Let us consider a disjoint division of the interval, $\Omega$, over
which the distribution $p(x)$ is defined. Each of these subintervals
have an index $i$ running from $1$ to $N$ and a size of $\omega_i$,
such that $\sum_i\omega_i=\Omega$. Then let us define a characteristic
function $\chi_i(x)$ 
\beq
\chi_i(x)=\left\{
\begin{aligned}
1&\qquad x\in\omega_i\\
0&\qquad \mbox{otherwise}
\end{aligned}\right.
\eeq
These functions are orthogonal,
\beq
\int dx\,\chi_i(x)\chi_j(x)=\omega_i\delta_{ij},
\label{ort}
\eeq
where $\delta_{ij}$ is the Kronecker delta.
The coarse grained value of the distribution $p(x)$ in interval $i$
is $p(x)\omega_i=Q_i$, if $x\in\omega_i$, more precisely 
\beq
Q_i=\int dx\,p(x)\chi_i(x).
\eeq 
This way our coarse grained approximation to the density function is
\beq
\tilde{p}(x)=\sum_i\frac{Q_i}{\omega_i}\chi_i(x),
\label{cga}
\eeq
that obviously satisfies the normalization condition
\beq
\int dx\,\tilde{p}(x)=\sum_iQ_i=1.
\eeq
On the other hand the integral of the square of this function using
Eqs. (\ref{ort}) and (\ref{cga}) is
\beq
\int dx\,[\tilde{p}(x)]^2=\sum_i\frac{Q_i^2}{\omega_i}
=\frac{1}{\omega}\sum_iQ_i^2.
\label{qq1}
\eeq
For sake of simplicity we have used (and will use from now on) an 
equipartition, $\omega_i=\omega=\Omega/N$. Now let us calculate the 
participation number, $D$, and entropy, $S$, using Eqs.~(\ref{DS}) 
from the discrete sums over the probabilities $Q_i$. First it is clear 
from Eq.~(\ref{qq1}) that
\beq
-\ln D=\ln\sum_iQ_i^2=\ln\int dx\,[\tilde{p}(x)]^2+\ln\omega,
\label{D2}
\eeq
Similar procedure leads to 
\beq
S=-\sum_iQ_i\ln Q_i=-\int dx\,\tilde{p}(x)\ln\tilde{p}(x)
-\ln\omega,
\label{S2}
\eeq
The appearance of the term $\ln\omega$ in both of these equations shows 
another type of divergence originating from the subdivision of the interval 
$\Omega$, since the limit of $N\to\infty$ corresponds to $\omega\to
0$. Therefore a naive application of these quantities may encounter
severe conceptual and also numerical difficulties depending on the
value of $\omega$. On the other hand we may conclude once again that 
the parameters $q$ and $S_{str}$ are free from this type of divergence, 
as well as using Eqs. (\ref{D&S}), (\ref{sstr}), (\ref{ff})
\begin{subequations}
\bea
-\ln q&=&\ln\left[\Omega\int dx\,[\tilde{p}(x)]^2\right],
\label{q2}\\
S_{str}&=&-\int dx\,\tilde{p}(x)\ln\tilde{p}(x)
        +\ln\int dx\,[\tilde{p}(x)]^2.
\label{s2}
\eea
\label{qs2}
\end{subequations}
This way we have shown how to apply the formalism developed for
discrete sums for the problem of continuous distributions. 

\section{Phase space representation of quantum states}

One of the most well-known phase space distributions that is widely
applied in statistical physics is the Wigner--function associated with
the quantum state \cite{HWL} $\psi(x)$
\beq
W(x,p)=\int dx'\,e^{-ipx'/\hbar}
\psi^*\left(x-\frac{x'}{2}\right)\psi\left(x+\frac{x'}{2}\right).
\label{wig}
\eeq
From now on $p$ denotes momentum and for sake of simplicity we
consider only one degree of freedom resulting in a two dimensional
phase space of $(x,p)$.

It is known that $W(x,p)$ is bilinear, real and for a complete
orthonormal set of $\psi_i$ functions the corresponding Wigner
transforms also form a complete orthonormal set \cite{HWL}. The marginal
distributions of $W(x,p)$ have an important physical meaning
\begin{subequations}
\label{wmarg}
\bea
\int dp\,W(x,p)&=&|\psi(x)|^2\label{psix},\\
\int dx\,W(x,p)&=&|\phi(p)|^2\label{psik},
\eea
\end{subequations}
where $\phi(p)$ denotes the Fourier transform of $\psi(x)$
\beq
\phi(p)=\frac{1}{\sqrt{2\pi\hbar}}\int dx\,\psi(x)e^{-ipx/\hbar}.
\eeq
The only major disadvantage of the Wigner distribution is that is may
attain negative values albeit in a region of phase space smaller than
$\hbar$. It has been shown already by Wigner \cite{wigner} that there
exists no phase space distribution that would have all the above
properties and besides that to be nonnegative.

Another very popular phase space distribution is the Husimi function
\cite{HWL} that is obtained as the Gaussian smearing of the Wigner
function, $W(x,p)$
\bea
\rho(x,p)&=&\int dx'dp'\,W(x',p')\times\nonumber \\
&&\exp\left(-\frac{(x-x')^2}{2\sigma_x^2}-\frac{(p-p')^2}{2\sigma_p^2}\right)
\label{husimi}
\eea
where $\sigma_x\sigma_p=\hbar/2$ ensures minimum uncertainty. 

It is known \cite{HWL} that the Husimi function in bilinear, 
real valued and nonnegative but unfortunately produces an over-complete 
set of functions and moreover the marginal distributions do not have such a
transparent meaning as in Eq. (\ref{wmarg}). In fact the latter point
can be refined. Let us calculate these marginal distributions and will 
find that indeed they are the Gaussian smeared distributions in $x-$ and 
$p-$representations, respectively \cite{ball}. In order to show this we write 
(\ref{husimi}) in the form of a convolution of $W(x,p)$ with two Gaussian 
functions, $g_{\sigma_x}(x)$ and $g_{\sigma_p}(p)$ of the form of Eq. 
(\ref{gauss})
\beq
\rho(x,p)=\int dx'dp'\,g_{\sigma_x}(x-x')g_{\sigma_p}(p-p')W(x',p').
\eeq
Then similarly to Eq. (\ref{wmarg})
\begin{subequations}
\label{hmarg}
\bea
\int dp\,\rho(x,p)&=&\zeta(x)\label{spx},\\
\int dx\,\rho(x,p)&=&\eta(p)\label{spk},
\eea
\end{subequations}
where the marginal distributions are nothing else but smeared 
distribution obtained from quantum state, $\psi(x)$ and $\phi(p)$,
respectively
\begin{subequations}
\label{smear}
\bea
\zeta(x)&=&\int dx'\,g_{\sigma_x}(x-x')|\psi(x')|^2
\label{sx},\\
\eta (p)&=&\int dp'\,g_{\sigma_p}(p-p')|\phi(p')|^2
\label{sk},
\eea
\end{subequations}
It is clear from the definition of the Husimi distribution that it 
is normalized as
\beq
\int dxdp\,\rho(x,p)=1,
\label{Hnorm}
\eeq
therefore the smeared $x-$representation of $\psi(x)$, $\zeta(x)$ and
the smeared $p$-representation of $\phi(p)$, $\eta(p)$ are normalized
as
\beq
\int dx\,\zeta(x)=\int dp\,\eta(p)=1.
\eeq

To complete this section we mention that the Husimi representation of 
a quantum state $\psi(x)$ is nothing else but its projection onto
(i.e. the overlap with) a coherent state with minimal uncertainty 
\cite{wobst,qfunc,Korsch,Karol01} $\beta_{x,p}(x')$ 
\beq
\rho_{\psi}(x,p)=|\langle\beta(x,p)|\psi\rangle|^2=
\left|\int dx'\beta_{x,p}^*(x')\psi(x')\right|^2
\label{hus1}
\eeq
the coherent state is a Gaussian centered around the phase space point
$(x,p)$
\beq
\beta_{x,p}(x')=\left(\frac{1}{2\pi\sigma^2}\right)^{1/4}
\exp\left(-\frac{(x'-x)^2}{4\sigma^2}+ipx'/\hbar\right).
\label{hus2}
\eeq

\section{R\'enyi entropies of phase space distributions}

In this section we describe how to characterize localization or
ergodicity using the ingredients explained in the previous sections: 
($i$) Husimi representation of the quantum states $\psi$ and ($ii$) 
the R\'enyi entropies, especially their differences. 

First of all let us introduce the R\'enyi entropies of the Husimi
function. In analogy with the definition (\ref{renyi2})
\beq
R_m=\frac{1}{1-m}\ln \int\frac{dxdp}{h}[h\rho(x,p)]^m
\label{renyicont}
\eeq
which now contains the arbitrary parameter $h$, that naturally behaves
as the minimum possible volume provided by the Heisenberg uncertainty
principle, i.e. it should be chosen as the Planck's constant. We have
to note that each $R_m$ contains $\ln h$ additively, that diverges in
the classical limit $h\to 0$ but drops out when differences of the
entropies are taken. Definition (\ref{renyicont}) for a compact phase 
space of volume $\Omega$, for instance, provides for the special case, 
$R_0=\ln (\Omega/h)$, i.e. it measures the size of the full phase 
space in units of $h$. Furthermore
\bea
\label{r12}
R_1=S&=&-\int dxdp\,\rho(x,p)\ln[h\rho(x,p)]\label{r1}\\
R_2&=&-\ln\left(h\int dxdp\, [\rho(x,p)]^2\right). \label{r2}
\eea
These are in accordance with Boltzmann's original definition, since 
for a distribution that is constant over a volume $\Gamma\leq\Omega$ 
and zero otherwise we obtain
\beq
S=\ln(\Gamma/h), \qquad\mbox{and}\qquad q=\Gamma/\Omega,
\eeq
i.e.  $S$ measures the size of phase space, $\Gamma$, where $\rho$ is
nonzero in units of $h$ and $q$ measures the portion of phase space 
where $\rho$ is different from zero.

In order to relate the entropy of the total Husimi function to that of
the marginal distributions
let us invoke an important relation that has been proven for the
Shannon entropy, $S$. Consider a distribution $\rho(x,p)$ which in our
case is the Husimi distribution, see Eq. (\ref{husimi}) or Eq.
(\ref{hus1}). Its information content, or Shannon entropy \cite{wehrl}
(\ref{r1})
can be related to the Shannon entropy of the marginal distributions,
$S[\zeta]$ and $S[\eta]$ defined in (\ref{DS_s}) for $\zeta(x)$ and 
$\eta(p)$ (\ref{smear}), respectively. Let us note that the Husimi 
distribution can be written in the form
\beq
\rho(x,p)=\zeta(x)\eta(p)+\delta(x,p),
\label{Hfact}
\eeq
where 
\beq
\int dx\,\delta(x,p)=\int dp\,\delta(x,p)=0.
\label{dcond}
\eeq
The Shannon entropy then obbeys \cite{AH} the following relation 
\beq
S[\rho]+\ln h=S[\zeta]+S[\eta]+\delta S
\label{sdecomp}
\eeq
where $\delta S<0$. Equality is achieved if $\delta(x,k)=0$ everywhere.
This statement can be generalized to R\'enyi entropies where 
\beq
R_m[\rho]+\ln h=R_m[\zeta]+R_m[\eta]+\delta R_m
\label{rdecomp}
\eeq
with $\delta R_m<0$. Unfortunately there is no general law for the
size of the $\delta R_m$, however, for the differences of R\'enyi
entropies it may become only a small correction if $\delta(x,p)\ll
\zeta(x)\eta(p)$. Furthermore for the parameters $-\ln q=R_2-R_0$ and 
$S_{str}=R_1-R_2$ we should have corrections of $\delta R_2-\delta R_0$
and $\delta R_1-\delta R_2$. These differences especially after averaging 
over several wave functions can be neglected, thereofore we arrive at the
following approximate relation for the average values of $-\ln q$ and 
$S_{str}$
\begin{subequations}
\label{addit}
\bea
-\ln q[\rho]&\approx &-\ln q[\zeta]-\ln q[\eta]\label{addq}\\
S_{str}[\rho]&\approx &S_{str}[\zeta]+S_{str}[\eta]\label{adds}
\eea
\end{subequations}
Such relations reduce the calculations considerably as there
would be no need to calculate the Husimi functions themselves and then
calculate the R\'enyi entropies thereof. In a numerical application we
will show below that indeed these relations do hold with small error.

We have to stress that these approximate relations may hold apart from the
trivial case of the wave packet (presented next) for the average 
properties of a suitably chosen set of states and most importantly in 
the case of the lack of underlying classical dynamics. These limitations 
reduce its applicability for the investigation of the states of disordered 
systems, which is nevertheless the main aim of our study.

As a simple example we elaborate the case of a distribution that is a 
Gaussian in both coordinates $x$ and $p$
\beq
\rho(x,p)=\frac{1}{\pi\hbar}
\frac{\exp\left(-x^2/2\sigma^2-2(\sigma p/\hbar)^2\right)}
{\Phi(\alpha/\sqrt{8})\Phi(\sqrt{2}\pi\beta)},
\label{ro}
\eeq
and normalized over the phase space bounded as $-L/2\leq x\leq L/2$
and $-\pi\hbar/a\leq p\leq \pi\hbar/a$, where two cutoff length
scales, $L$ and $a$, have been introduced. Therefore the volume of the 
phase space will be $\Omega=hL/a=\gamma h$. The ratios of the cutoff 
scales to the spreading width $\sigma$ yield the two dimensionless 
parameters in (\ref{ro}), $\alpha=L/\sigma$ and $\beta=\sigma/a$. 
In terms of these quantities we obtain the relation 
$\gamma =\alpha\beta$, which counts the number of cells of size $h$ in 
phase space.

Since in (\ref{ro}) $\sigma_x=\sigma$ and $\sigma_p=\hbar/2\sigma$, the
uncertainty relation, $\sigma_x\sigma_p=\hbar/2$, is fulfilled. This
$\rho(x,p)$ is in fact the Husimi distribution of a real space Gaussian 
wave packet and it is a product of the limit distributions as given in
(\ref{Hfact}) with $\delta(x,p)=0$.
Consequently the $q$ and $S_{str}$ values of the corresponding limit
distributions, $\zeta(x)$ and $\eta(p)$, obey the additivity property 
(\ref{addit}) exactly. Distribution $\zeta(x)$, for instance, is obtained 
using (\ref{spx}) and yields Eq.~(\ref{ngauss}). Its $q[\zeta]$ and
$S_{str}[\zeta]$ parameters are given in Eqs.~(\ref{qxi}) and (\ref{sxi}). 
A straightforward calculation yields $q[\eta]$ and $S_{str}[\eta]$, as
well.

Putting the Husimi distribution (\ref{ro}) into (\ref{renyicont}) we
find
\bea
R_m(\alpha,\beta)&=&\frac{\ln m}{m-1}-\ln 2 + \nonumber \\
&&
\frac{1}{m-1}\ln\left\{\frac{\Phi(\alpha\sqrt{m/8})\Phi(\beta\pi\sqrt{2m})}
{\left[\Phi(\alpha/\sqrt{8})\Phi(\beta\pi\sqrt{2})\right]^m}\right\}
\label{rewp}
\eea
This expression correctly yields $R_0=\ln(\alpha\beta)=\ln(\gamma)$,
the log of the number of cells of size $h$. Through relations
(\ref{sq_renyi}) parameters $q$ and $S_{str}$ can be obtained.
Equivalently using the definition (\ref{ro}) and (\ref{qs2}) 
generalized for the case of the Husimi distribution like in 
(\ref{renyicont}) and (\ref{r1}) we obtain 
\begin{subequations}
\bea
q(\alpha,\beta)&=&\frac{[\Phi(\alpha/\sqrt{8})\Phi(\beta\sqrt{2}\pi)]^2}
	    {\alpha\beta\Phi(\alpha/2)\Phi(2\pi\beta)}\label{qc}\\
S_{str}(\alpha,\beta)&=&1-\ln(2)\nonumber \\
  &&-\frac{\sqrt{2\pi}\beta e^{-2(\pi\beta)^2}}{\Phi(\sqrt{2}\pi\beta)}
    -\frac{\alpha e^{-\alpha^2/8}}{\sqrt{8\pi}\Phi(\alpha/\sqrt{8})}
    \nonumber
	    \\
  &&+\ln\left[
     \frac{\Phi(2\pi\beta)\Phi(\alpha/2)}
          {\Phi(\sqrt{2}\pi\beta)\Phi(\alpha/\sqrt{8})}
	\right]\label{sc}
\eea
\label{qsc}
\end{subequations}
There are a number of remarkable features of Eqs. (\ref{rewp}) and
(\ref{sc}). None of them contain Planck's constant, $h$, explicitly, 
however, the scaling variables $\alpha$ and $\beta$ cannot be chosen
independently as their product is just the number of cells of size $h$.
Therefore let us keep $\alpha$ as a running variable and parametrize the 
functions with $\gamma$. Furthermore, let us note that by keeping
$\gamma$ fixed the $q(\alpha)$ and $S_{str}(\alpha)$ functions are
symmetrical about $\alpha_0=\sqrt{4\pi\gamma}$ on a logarithmic scale.
Therefore rewriting (\ref{qsc}) in the variable $t=\alpha/\alpha_0$ we
obtain
\bea
q^{(\gamma)}(t)&=&\frac{1}{\gamma}
                \frac{\left[\Phi(c_1/t)\Phi(c_1 t)\right]^2}
		     {\Phi(c_2/t)\Phi(c_2 t)}\nonumber \\
S_{str}^{(\gamma)}(t)&=&1-\ln(2)\nonumber\\
  &&-\sqrt{ \frac{\gamma}{2} }
    \left\{\frac{te^{-(c_1t)^2}}{\Phi(c_1t)}
          +\frac{e^{-(c_1/t)^2}}{t\Phi(c_1/t)}
    \right\}\nonumber \\
  &&+\ln\left[\frac{\Phi(c_2/t)\Phi(c_2 t)}{\Phi(c_1/t)\Phi(c_1 t)}
        \right]
\label{sc2}
\eea
\begin{figure}
\includegraphics[width=3in]{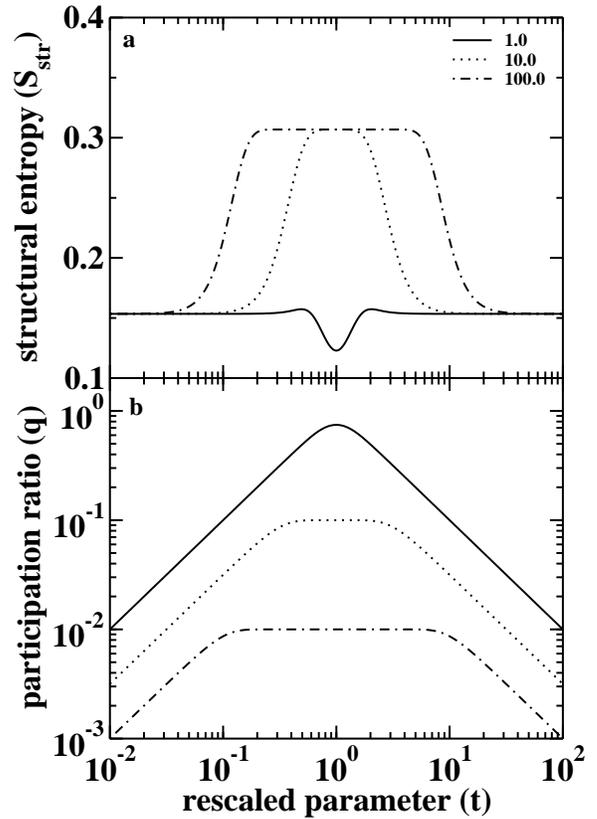}
\caption{\label{coh}
The parameters $S_{str}$ (\textbf{a}, semi-log plot) and
$q$ (\textbf{b}, log-log plot) for a Gaussian wave packet as a function 
of the parameter $t=\alpha/\alpha_0$, where $\alpha=L/\sigma$, $L$ is 
the system size and $\sigma$ is the spreading of the Gaussian.
$\alpha_0=\sqrt{4\pi\gamma}$. The different curves are parametrized 
according to $\gamma$, the number of cells within $h$.
}
\end{figure}
where $c_1=\sqrt{\pi\gamma/2}$ and $c_2=\sqrt{2}c_1$.
These functions are plotted in Figure \ref{coh}. 
At $t=1$ function $q(t)$ is maximal and its value decreases with
$\gamma$ as $q(1)\propto\gamma^{-1}$. 
Only the physically relevant, $\gamma\geq 1$ are
plotted. The participation ratio has some very nice, simple behavior,
$q^{(\gamma)}(t)\to (t\sqrt{\gamma})^{-1}$ for $t\to\infty$ and
$q^{(\gamma)}(t)\to t/\sqrt{\gamma}$ for $t\to 0$. On the other hand
in the same limits $S_{str}\to S_{str}^{G}$ independently from
$\gamma$ showing that these limits correspond to one dimensional
Gaussians in $x$($p$--)directions for $t\to\infty(t\to 0)$. It can
also be viewed as if a squeezing parameter made the distribution
more coordinate-like (momentum-like) \cite{korsch}.
For intermediate values of $t$, i.e. if $1/\sqrt{\gamma}<t<\sqrt{\gamma}$
with $\gamma\gg 1$ the participation ratio is, $q\approx\gamma^{-1}$ and 
$S_{str}\approx 2S_{str}^G$ indicating that this distribution is a Gaussian 
in both dimensions, $x$ and $p$. This is the regime where the 
Husimi--function is a Gaussian in both directions and therefore shows a 
two-dimensional character. Both curves, $q$ anf $S_{str}$ are 
symmetrical about $t=1$ on a logarithmic scale of $t$,
which is a direct consequence of the geometry of the phase space. 

\section{Application to disordered systems}

Now we calculate Husimi functions of the eigenstates of a disordered 
one dimensional system. To be more precise we use a tight binding model 
\cite{KM}
\beq
H=\sum_n \varepsilon_n|n\rangle\langle n|
 +V\sum_n\left(|n\rangle\langle n+1|+|n+1\rangle\langle n|\right)
\label{hamil} 
\eeq
where $V=1$ is set as the unit of energy and $\varepsilon_n$ are
random numbers distributed uniformly over the interval
$[-W/2,\dots,W/2]$, where $W$ characterizes the strength of disorder.
Such a model has been investigated in phase space in
Refs.~\onlinecite{dietmar} and \onlinecite{wobst}. The Husimi functions 
are calculated using Eqs.~(\ref{hus1}) and (\ref{hus2}) from the 
eigenstates of (\ref{hamil}). The participation ratio, $q$, and the 
structural entropy, $S_{str}$, are calculated according to Eq. 
(\ref{qs2}). We also calculated the Fourier transforms of the 
eigenstates and obtained smeared distributions according to Eq. 
(\ref{smear}) both in real and Fourier space. Periodic boundary 
conditions were considered using $L=512$. The phase space extends over 
$-L/2\leq x\leq L/2$ and $-\pi<p\leq\pi$, its area is $\Omega=2\pi L$ 
($\hbar=1$ and the lower cut-off scale, the lattice spacing is set to
unity, $a=1$). Averaging is done over the middle half of the band. 
In fact, as pointed out by Ref. \onlinecite{wobst}, as well, there 
is no need to average over many realizations of the disordered potential. 
\begin{figure}
\includegraphics[width=3in]{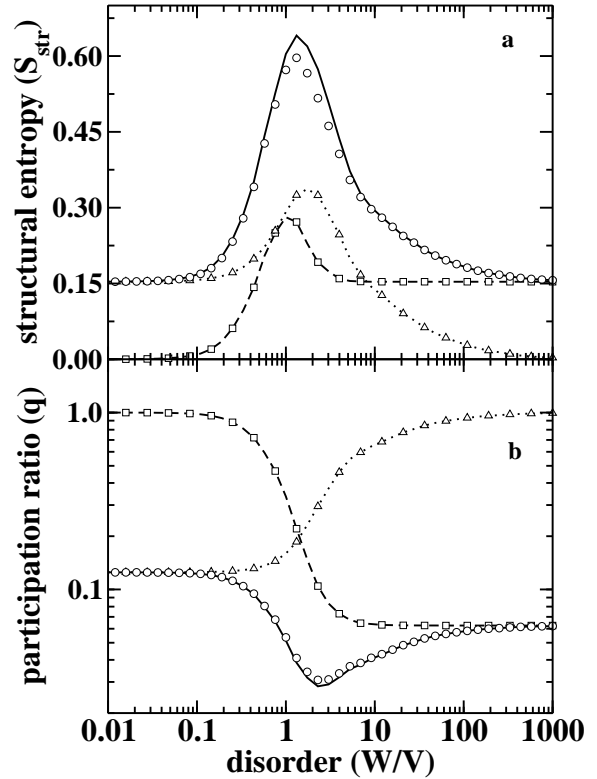}
\caption{\label{f2}
Structural entropy, $S_{str}$ (\textbf{a}, semi-log plot) and
participation ratio, $q$ (\textbf{b}, log-log plot)  
as a function of disorder, $W$ in units of $V$, for a 
one dimensional Anderson model with $L=512$. The squares stand for 
the states smeared in $x$-representation, the triangles for the states 
smeared in $p$-representation, the circles are calculated according to 
Eqs. (\protect\ref{addit}). The dotted lines are simply guides for the
eyes. The solid curve corresponds to the values of $S_{str}$ and $q$
for the states in Husimi representation. The circles and the solid 
curve differ only a little.}
\end{figure}

When the full Husimi distributions of all states are calculated the 
computational time grows with $L^4$. However, using the approximate
relation of Eq. (\ref{addit}), it reduces to roughly $L^3$. This is
obviously a considerable gain and is comparable to the one achieved in 
Refs.~\onlinecite{sugita,wobst}.

The results are reported in Fig. {\ref{f2}. Here we have plotted the
behavior of parameters (\ref{qs2}) versus disorder strength, $W$, and 
compared to the approximate values obtained using (\ref{addit}). 
The region where the most important variations of
$q$ and $S_{str}$ take place is $W_1<W<W_2$, where the localization 
length matches the systems size \cite{KM,vinew}, $\lambda\approx L$, 
i.e. $W_1\approx \sqrt{105/L}=0.453$ or its inverse approaches the 
size of phase space in $p$--direction, $2\pi\lambda\approx 1$, i.e.
$W_2\approx\sqrt{2\pi 105}=25.68$

Let us analyze the expectations for the limiting cases of $W\to 0$ and
$W\to\infty$. It is easy to see that the eigenstates for
vanishing disorder are plane waves whose Fourier transform is a
Dirac--delta (in fact two, due to the symmetry of the $-p$ and $p$
states). In a `smeared' representation we obtain two Gaussians in
$p$-representation. Using Eq. (\ref{hus2}) and (\ref{q2}) we obtain
that $q=2/\sqrt{L}$ for the states both in $p$- and Husimi representation. 
Due to the Gaussian smearing in this limit the structural entropy
attains its value of $S_{str}^G$ as given in (\ref{ggg}). The other
limit of $W\to\infty$ is very similar. In that case the eigenstate in
$x$-representation has a Dirac--delta character that is smeared
to a Gaussian. This results in $q=1/\sqrt{L}$ and again a value of
$S_{str}(W)\to S_{str}^G$. All these limiting cases are recovered in
Fig. \ref{f2}. The figure shows that the approximation (\ref{addit}) 
works very well. It is clear that the $x$-, $p$- and Husimi 
representations are therefore linked very simply.

\section{Concluding remarks}

In this paper we have presented some important results concerning
the applicability of R\'enyi entropies for the characterization of
localization or ergodicity in phase space using the Husimi
representation of the quantum states. In fact it has been shown that the
differences of R\'enyi entropies are free from those divergences that
would naturally arise due to their application on continuous distributions.

The marginal distributions of the Husimi function are pointed out to
have important properties and simple connection to the states in $x$-
and $p$-representations.

We have also shown numerically that for disordered systems the
limiting distributions of the Husimi function provide most of the
information that is needed to  describe the Husimi functions
themselves. Figure \ref{f2} provides a good demonstration of the 
duality between the $x$- and $p$-representations transparently. A
detailed study over the Anderson model in one dimension and the 
Harper model \cite{vinew} are left for forthcoming publications.

\begin{acknowledgments}
One of the authors (I.V.) acknowledges enlightening discussions with
B. Eckhardt, P. H\"anggi, G-L. Ingold, and A. Wobst.
Work was supported by the Alexander von Humboldt Foundation, the 
Hungarian Research Fund (OTKA) under T032116, T034832, and T042981.
\end{acknowledgments}

\end{document}